\documentstyle[12pt]{article}

\def\Journal#1#2#3#4{{#1} {\bf #2}, #3 (#4)}

\def\abstracts#1{
\begin{center}
{\begin{minipage}{4.2truein}
                 \normalsize
                 \parindent=0pt #1\par
                 \end{minipage}}\end{center}
                 \vskip 2em \par}

\def\half{\frac{1}{2}}

\def\beq{\begin{equation}}
\def\eeq{\end{equation}}
\def\bea{\begin{eqnarray}}
\def\eea{\end{eqnarray}}
\def\xmu{x^{\mu}}
\def\xnu{x^{\nu}}
\def\eps{\varepsilon}
\newcommand{\IR}{\mbox{I\kern -0.25em R}}
\def\ie{{\it i.e.\ }}

\title{\noindent {\small\hfill UTAS-PHYS-98-19} \\ 
{\vskip 1.0cm}
The $D(2,1;\alpha)$ Particle\footnote{to appear in the {\it Proceedings for the 22nd International Colloquium on Group Theoretical Methods in Physics, International Press, Cambridge, MA, USA}}}

\author{S.\ P.\ Corney, P.\ D.\ Jarvis and D.\ S.\ McAnally\footnote{Department of Mathematics, University of Queensland St Lucia, Brisbane, Queensland, Brisbane, Australia, 4072} \\
School of Mathematics and Physics, University of Tasmania \\
GPO Box 252-21, Hobart, Tasmania, Australia, 7001 \\
Email: Stuart.Corney@utas.edu.au}

\begin{document}

\maketitle

\abstracts{The exceptional superalgebra $D(2,1;\alpha)$ has been
classified as a candidate conformal supersymmetry algebra in two
dimensions.  We 
propose an alternative interpretation of it as extended BFV-BRST 
quantisation  superalgebras in $2D$ ($D(2,1;1) \simeq osp(2,2|2)$).  A 
superfield
realization is presented wherein the standard extended 
phase space coordinates can be identified. The physical states are
studied via the cohomology of the
BRST operator. 
}

The BFV-BRST quantisation of relativistic systems provides a cohomological resolution of irreducible unitary representations of space-time symmetries. Moreover these unirreps appear to be associated with constructions of $iosp(d,2|2)$\cite{jar}, for relativistic particles in flat spacetime, however, in this work we do not need or use translations\cite{bar}. Here we follow an algebraic approach, and so need to develop a classification of admissible `quantisation superalgebras' in various dimensions. Some examples of such algebras\cite{gun} are $D(2,1;\alpha)$ in $D = 1+1$ (note that as $\alpha =1$ corresponds to $osp(2,2|2)$), in $D =2+1$ we have $osp(3,2|2)$ which corresponds to anti de Sitter symmetry (which may be relevant to anyon quantisation), and for $D = 3+1$ we get conformal symmetry of $4D$ spacetime and super unitary superalgebras.

As is well known \cite{gov,hen} the BFV canonical quantisation of canonical constrained Hamiltonian systems \cite{BFV} uses an extended phase space description in which, associated with each first class constraint $\phi_{(a)}$ a pair of conjugate variables $\eta^{(a)},\rho_{(a)}$is introduced. These ghost variables have Grassman parity opposite to that of the constraint. Including Lagrange multipliers $\lambda$ as additional dynamical variables, leads to their vanishing canonical momenta ($\pi_\lambda$) constituting an auxiliary, doubled set of first class constraints. The action for the scalar particle can be written (in $2^{nd}$ order formulation):
\beq
S = m \int_{\tau_{i}}^{\tau_{f}} d\!\tau \sqrt{\frac{d\xmu}{d\tau}\frac{d\xnu}{d\tau}}.
\eeq
The constraints are the mass-shell condition $\phi_1 = p^\mu p_\mu - m^2 = 0$, as well as the aforementioned vanishing conjugate momentum of the Lagrange multiplier, which shall be slightly modified shortly. The extended action is invariant under the infinitesimal guage transformations $\delta \lambda = \dot{\eps}, \;\delta \xmu = \{\xmu,\eps\Phi\} = 2 \eps p^\mu, \; \delta p^\mu = \{p^\mu, \eps \Phi \} =0.$ In order to solve for the action it is necessary to choose a particular gauge fixing condition. From the transformation of the einbein (which arises explicitly in the first order formulation) under world-line diffeomorphisms, we have $e'(\tau')d\tau' = d\tau e(\tau)$. In the infinitesimal case $\tau = \tau' + \eps(\tau')/e(\tau)$,  we have $\delta e = e'(\tau) - e(\tau) = \dot{\eps}$. Thus we identify $\lambda(\tau)$ with $e(\tau)$.

Two restrictions are necessary so as to arrive at the particle quantisation corresponding with the superalgebraic prescription. Firstly we must limit the quantisation of $\lambda$ to the half-line (say ${\IR}^+$) \cite{gov}, or equivalently we fix the gauge so as to be with respect to gauge transformations in one of the connected components of the group, \ie the identity class or the orientation reversing class. Secondly, we take $\phi_{2} = \lambda \pi_{\lambda}$ as the second constraint (rather than $\phi_{2} =\pi_{\lambda}$ as used in the standard construction)\footnote{We thank J Klauder and J Govaerts for pointing out
possible difficulties with regularity of the constraints in this case}. 

The standard BRST operator can now be written $\Omega = \eta^{(1)}\lambda\pi_{\lambda} + \eta^{(2)}(p^\mu p_\mu - m^2)$. The gauge fixed Hamiltonian is $H = \{ {\mathcal F},\Omega \}$, which for the choice of guage fixing funtion ${\mathcal F} = \half \lambda \rho_{(2)}$ yields $H = - \half \lambda (\eta^{(1)} \rho_{(2)} + p^{\mu} p_{\mu} - m^2)$.

The system can be quantised by introducing the standard Schr\"{o}dinger representation. It is at this point convenient to change variables and rescale as follows: define $p_+ = \lambda^{-1}$ and $x_- = \lambda\pi_{\lambda}$ at the classical level, and implement a canonical rescaling $\eta \rightarrow \lambda\eta, \; \rho \rightarrow \lambda^{-1}\rho$ as well as $\pi_{\lambda} \rightarrow \pi_{\lambda} - (\eta^{(1)} \rho_{(1)}- \eta^{(2)} \rho_{(2)})$. We now introduce the quantum operators $X^\mu,\;P_\nu$ corresponding to the co-ordinates $x^\mu\;p_\nu$, along with $P_+$ and $X_-$ acting on suitable sets of wavefunctions on the half line. The hermitean ghosts $\eta^{(a)}, \rho_{(b)}$ are represented on a 4-dimensional indefinite inner product space $|\sigma\sigma'\rangle, \sigma,\sigma' = \pm,$ on which the combinations $Q_{1,2} = (\eta^{(1)} \pm \rho_{(2)})/\sqrt{2}$ satisfy $[Q_\alpha,X_\beta] = \eps_{\alpha\beta}$ (see \cite{jar} for details). Finally, with
\beq
P_- = H = -\half P_+^{-1}(P^\mu P_\mu + Q^\alpha Q_\alpha -m^2)
\eeq
we find that operators $J_{AB}, A = \mu, \pm ({\rm even}), \alpha ({\rm odd}), B = \nu,\pm,\beta \; etc.$ defined by bilinear antisymmetrised combinations of $X_\mu$ (or $X_\alpha$) and $P_\nu$ (or $Q_\beta$) close on the inhomogeneous orthosymplectic superalgebra $isop(d,2/2)$,
\beq
{[}J_{AB},J_{CD}{]} = g_{CB} J_{AD} - [AB] g_{CA} J_{BD }
 - [CD] g_{DB} J_{AC} + [AB] [CD] g_{DA} J_{BC}
\eeq
in a light cone basis where $x_\pm = \frac{1}{\sqrt{2}}(x_d\pm x_{d+1})$, the metric is $g_{\mu\nu} = \eta_{\mu\nu}, \;g_{+-} = g_{-+} =1,\;g_{\alpha\beta} = \eps_{\alpha\beta}$, and $[A,B] \equiv (-1)^{ab}$ etc..

For the $D(2,1;\alpha)$ particle we start with no corresponding model, and so have only the algebraic structure as a guide. We regard $D(2,1;\alpha)$ as a generalisation of $osp(d,2|2)$ and find a superfield realisation which is equivalent to the case for a relativistic scalar particle \cite{jar} for $\alpha = 1$ and $d=2$. In the generic $d$ case we define the homogeneous manifold ${\mathcal M} = OSp(d-1,2|2) / G_0$, where $OSp(d-1,1|2) = \langle J_{\mu\nu},L_{\mu\alpha},K_{\alpha\beta} \rangle$ and ${\mathcal N} = \langle J_{\mu -}, L_{\alpha -} \rangle$. The stability group is therefore $G_0 = OSp(d-1,1|2) \wedge {\mathcal N}$. For one parameter subgroups $g(t)$ with generator $A$,  the standard superfield realisation leads to generators, $\phi$ acting on ${\mathcal M}$, defined by
\beq
\hat{A}\phi(x) = \left(\frac{d}{dt}\Phi(g(t)^{-1}x) \right)_{t=0},
\eeq
where ${\mathcal M} \ni x = (q^\mu ,\eta^\alpha , \phi) \leftrightarrow \exp (q^\mu J_{\mu+} + \eta^\alpha L_{\alpha+}) \exp(\phi J_{+-})G_0$, represents the coset.

In the case of the $D(2,1;\alpha)$ superalgebra, we define generators $J_{\mu\nu} = \eps_{\mu\nu}J, \tilde{J} = J + aJ_{+-}$ and $\tilde{J}_{+-} = J_{+-} + aJ$, where $a = \frac{1-\alpha}{1+\alpha}$, and the remaining generators are as in the standard superfield case. The nonzero anticommutators which differ from the $osp(d,2|2)$ case are
\bea
\{L_{\mu\alpha},L_{\nu\beta} \} &=& \eps_{\alpha\beta}\eps_{\mu\nu}\tilde{J} - \eta_{\mu\nu}K_{\alpha\beta} \nonumber\\
\{ L_{\mu\alpha},L_{\beta\pm} \} &=& -\eps_{\alpha\beta}(J_{\mu\pm} \pm a \eps_\mu ^\nu J_{\nu\pm}), \\
\{ L_{\alpha\pm},L_{\beta\mp} \} &=& \pm \eps_{\alpha\beta} \tilde{J}_{+-} - K_{\alpha\beta}. \nonumber
\eea
We can then define generators $\hat{A}$ similar to the standard case, except now $OSp(1,1|2) = \langle \tilde{J},L_{\mu\alpha},K_{\alpha\beta} \rangle$, leading to $\tilde{G}_0$, and so ${\mathcal M} \ni x \equiv (q^\mu ,\eta^\alpha , \phi) \leftrightarrow \exp (q^\mu J_{\mu+} + \eta^\alpha L_{\alpha+}) \exp(\phi \tilde{J}_{+-})\tilde{G}_0$.

We can now explicitly write down the superfield realisation for $D(2,1;\alpha)$ as follows: let $p^\mu = \lambda^{-1}, \theta^{\alpha} = \lambda^{-1}\eta^{\alpha}, \lambda = e^{\phi} (\lambda > 0)$, therefore
\bea
J_{\mu +} &=& - \lambda^{-1} \frac{\partial}{\partial p^{\mu}}, \nonumber\\
L_{\alpha +} &=& - \lambda^{-1} \frac{\partial}{\partial \theta^{\alpha}},
\nonumber\\
L_{\alpha -} &=& \frac{1}{2} \lambda (p^{\nu} p_{\nu}
+ \theta^{\beta} \theta_{\beta}) \frac{\partial}{\partial \theta^{\alpha}}
- \theta_{\alpha} \lambda^2 \frac{\partial}{\partial \lambda}
- a \lambda \theta_{\alpha} p^{\mu} \varepsilon_{\mu}{}^{\nu}
\frac{\partial}{\partial p^{\nu}}, \nonumber\\
L_{\mu \alpha} &=& p_{\mu} \frac{\partial}{\partial \theta^{\alpha}}
- \theta_{\alpha} \frac{\partial}{\partial p^{\mu}} - a \theta_{\alpha}
\varepsilon_{\mu}{}^{\nu} \frac{\partial}{\partial p^{\nu}}, \nonumber\\
K_{\alpha \beta} &=& \theta_{\alpha} \frac{\partial}{\partial \theta^{\beta}}
+ \theta_{\beta} \frac{\partial}{\partial \theta^{\alpha}}, \\
J &=& - p^{\mu} \varepsilon_{\mu}{}^{\nu} \frac{\partial}{\partial p^{\nu}}
+ \frac{a}{1-a^2} \left(\lambda \frac{\partial}{\partial \lambda}
- p^{\mu} \frac{\partial}{\partial p^{\mu}} - \theta^{\alpha}
\frac{\partial}{\partial \theta^{\alpha}} \right), \nonumber\\
J_{+-} &=& - \lambda \frac{\partial}{\partial \lambda}
- \frac{a^2}{1-a^2} \left(\lambda \frac{\partial}{\partial \lambda}
- p^{\mu} \frac{\partial}{\partial p^{\mu}} - \theta^{\alpha}
\frac{\partial}{\partial \theta^{\alpha}} \right), \nonumber \\
J_{\mu -} &=& \frac{1}{2} \lambda\theta^{\alpha} \theta_{\alpha} 
\frac{\partial}{\partial p^{\mu}} 
+ \frac{1}{2} \lambda a \varepsilon_{\mu}{}^{\nu} \theta^{\alpha} 
\theta_{\alpha} \frac{\partial}{\partial p^{\nu}} 
+ \frac{1}{2} \lambda \varepsilon_{\mu \nu} \varepsilon_{\rho}{}^{\sigma} 
p^{\nu} p^{\rho} \frac{\partial}{\partial p^{\sigma}}  \nonumber\\
&&\mbox{\quad}- \lambda^2 p_{\mu} \frac{\partial}{\partial \lambda}
+ \frac{1}{2} \lambda p_{\mu} p^{\nu} \frac{\partial}{\partial p^{\nu}} 
\nonumber\\
&& \mbox{\quad}- \frac{\lambda a (a p_{\mu} + \varepsilon_{\mu \rho} p^{\rho})}
{1-a^2}
\left(\lambda \frac{\partial}{\partial \lambda} 
- p^{\nu} \frac{\partial}{\partial p^{\nu}} 
- \theta^{\alpha} \frac{\partial}{\partial \theta^{\alpha}} \right). \nonumber 
\eea
If we compare this realisation with that obtained for $osp(d,2|2)$\cite{jar} the similarities are evident (although the realisation of $J_{\mu -}$ requires some attention). In fact if we allow $\alpha \rightarrow 1$ (\ie $D(2,1;\alpha) \equiv osp(2,2|2)$) then these relations equal those obtained using the standard superfield.

The BRST operator for the $D(2,1;\alpha)$ model can be implemented by considering linearly independent spinors $(\chi_\alpha)$ and $(\chi'{}^{\alpha})$ with $\chi^\alpha \chi'_{\alpha} =1$, for example $(\chi^\alpha) = \frac{1}{\sqrt{2}}(1,1), \chi'{}^{\alpha} = \frac{1}{\sqrt{2}}(-1,1)$. Defining the ghost number operator $N = \chi^{\alpha}\chi'{}^{\beta}K_{\alpha\beta} \rightarrow (\chi \dot \theta)(\chi' \dot \partial) + (\chi' \dot \theta)(\chi \dot \partial)$, then for any spinor $V_{\gamma}$ we have $[N,\chi^\gamma V_\gamma ] = + \chi^{\gamma}V_{\gamma}$ and $[N,\chi'{}^{\gamma}V_{\gamma} ] = - \chi'{}^{\gamma} V_{\gamma}$. Thus we can define
\bea
\Omega \equiv \chi^{\alpha} L_{\alpha -}, & Q_{\alpha} \rightarrow \theta_{\alpha}, & X_{\alpha} \rightarrow \partial_{\alpha}, \nonumber \\
{\mathcal F} \equiv \chi'{}^{\alpha}Q_{\alpha}, & P_{\mu} \rightarrow p_{\mu}, & X_{\mu} \rightarrow i\frac{\partial}{\partial p^{\mu}}, \\
{\mathcal H} \equiv \{\Omega,{\mathcal F} \}. & & \nonumber
\eea

The physical states for our system can be stated explicitly by considering superfields of the form $\Phi(\xmu,p_{+},\tau,\theta^{\alpha}) = A + \theta^{\alpha} \psi_{\alpha} + \half\theta^{\alpha}\theta_{\alpha}B$, and imposing $\Omega \Phi = 0, \Phi \neq \Omega \Phi '$, thus we get
\bea
\Omega \Phi &=& \frac{1}{2p_+} p \cdot p (\chi \cdot \psi) 
+ \chi^{\alpha} \theta_{\alpha} \left( \frac{1}{2 p_+} B 
+ \left[ \frac{\partial}{\partial p_+} 
- \frac{a}{p_+} (p \varepsilon \partial_p) \right] A \right) \nonumber\\
&& + \frac{1}{2} \theta^{\alpha} \theta_{\alpha} 
\left(-\frac{\partial}{\partial p_+} + \frac{1}{p_+}(1+ap \varepsilon 
\partial_p) \right)(\chi \cdot \psi).
\eea
We note that in the Fourier space the constraints are solved by the physical states
$
\phi(x_R,x_L) \equiv \int (\chi \cdot \psi) e^{-i p \cdot x} dx,
$
satisfying
\beq
\frac{\partial}{\partial x_R} \frac{\partial}{\partial x_L} \phi = 0 \quad 
\left( \begin{array}{c} x_R = x^0+x^1\\x_L = x^0-x^1 \end{array} \right)
\eeq
and $\phi(x_R,x_L,p_+) = p_+ \varphi(p_+^a x_R,p_+^{-a} x_L)$ for some function $\varphi$. Moreover if ${\mathcal H}\Phi = 0$ on the physical states and we assume the Schr\"{o}dinger equation ${\mathcal H} = - i\frac{d}{d\tau}$, then these functions are independent of $\tau$.

The system we have constructed can be interpreted as the `quantisation' of a classical `$D(2,1;\alpha)$' particle. The $p_{+}-$dependence of physical wavefunctions provides indirect evidence that the model involves a more subtle implementation of diffeomorphisms than usual. Note that the two-dimensional case has the unique property that Lorentz invariance is not broken, and the metric $x^\mu x^\nu \eta_{\mu \nu} = x_{R}x_{L}$ is still a worldline scalar, if $x_{R}, x_{L}$ transform as densities, $x'_{R,L}(\tau') {d\tau'}^{\pm a} = x_{R,L}(\tau) {d\tau}^{\pm a}$. Corresponding covariant actions may be responsible (after gauge fixing) for the $p_{+}-$scaling behaviour\footnote{The gauge equivalence class of $\lambda$, or $e$, namely $\int_{\tau_{i}}^{\tau_{f}} e(\tau) d\,\tau$, is proportional to $\lambda$ in the present case $\dot{\lambda}=0$}. Finally, the superalgebraisation of BRST-BFV indicates a role for $(d,2|2)$ type superspaces. Future work will involve applying the $D(2,1;\alpha)$ model to the spinning particle\cite{tso}, as well as generalising to higher dimensions. 


\section*{Acknowledgments}
The authors wish to acknowledge the contribution provided by Ioannis Tsohantjis in the early stages of this work, as well as discussions with A J Bracken.


\end{document}